\documentclass[11pt]{article}
\usepackage{fullpage,epsf,amssymb,amsthm,amsfonts,amsmath,latexsym, graphicx,array,extarrows,mathtools,mathstyle,mathrsfs,slashed,subcaption,amsbsy}
\usepackage[normalem]{ulem}
\usepackage{authblk}
\usepackage{cite}
\usepackage{color}

\newcommand{\ud}{\mathrm{d}}
\newcommand{\uvec}[1]{\boldsymbol{#1}}

\title{\bf{Poincar\'e constraints on the gravitational form factors for massive states with arbitrary spin}}

\author[1]{Sabrina Cotogno\thanks{sabrina.cotogno@polytechnique.edu}} 
\author[1]{C\'{e}dric Lorc\'{e}\thanks{cedric.lorce@polytechnique.edu}}
\author[1]{Peter Lowdon\thanks{peter.lowdon@polytechnique.edu}}

\affil[1]{\small{\textit{CPHT, CNRS, Ecole polytechnique, IP Paris, F-91128 Palaiseau, France}}}

\date{}
\begin{document}

{\let\newpage\relax\maketitle}
\setcounter{page}{1}
\pagestyle{plain}

\abstract

\noindent
In this work we analyse the constraints imposed by Poincar\'e symmetry on the gravitational form factors appearing in the Lorentz decomposition of the energy-momentum tensor matrix elements for massive states with arbitrary spin. By adopting a distributional approach, we prove for the first time non-perturbatively that the zero momentum transfer limit of the leading two form factors in the decomposition are completely independent of the spin of the states. It turns out that these constraints arise due to the general Poincar\'e transformation and on-shell properties of the states, as opposed to the specific characteristics of the individual Poincar\'e generators themselves. By expressing these leading form factors in terms of generalised parton distributions, we subsequently derive the linear and angular momentum sum rules for states with arbitrary spin.

\newpage

\section{Introduction}

The matrix elements of local operators are of central importance in characterising the non-perturbative structure of any quantum field theory (QFT). In the case of the energy-momentum tensor (EMT) these matrix elements encode a wide variety of different phenomena, from the quantum corrections which arise in the gravitational motion of particles, to the distribution of mass and angular momentum within hadrons~\cite{Donoghue_Holstein_Garbrecht_Konstandin02,Polyakov:2002yz,Goeke:2007fp,Lorce:2017wkb,Polyakov:2018zvc,Lorce:2018egm}. Although there is a significant breadth of literature on these objects, most of these studies have chosen to focus on particular cases where the states have lower spin (generally spin 0, $\tfrac{1}{2}$, or 1~\cite{Polyakov:2002yz,Goeke:2007fp,Lorce:2017wkb,Polyakov:2018zvc,Lorce:2018egm,Kumano:2017lhr,Abidin:2008ku,Taneja:2011sy,Cosyn:2019aio,Polyakov:2019lbq}) or correspond to specific particles, as opposed to analysing the constraints imposed for arbitrary states. Whilst this approach has proven to be successful phenomenologically, it potentially risks obscuring the underlying properties governing these constraints, preventing one from separating model-specific and general QFT effects.\\

\noindent
An important feature of EMT matrix elements, like any other local matrix elements, is that they can be decomposed into a series of Lorentz structures. The coefficients of these terms, known as the gravitational form factors (GFFs), are constrained by the symmetry properties of the EMT, together with its conservation and the physical requirement that the states are on shell. Although this structure has been understood for many years, the subsequent form factor analyses have generally contained technical difficulties such as in the handling of boundary terms and the construction of well-defined normalisable states, leading to incorrect conclusions, as discussed in detail in~\cite{Bakker_Leader_Trueman04}. In~\cite{Lowdon_Chiu_Brodsky17} it was demonstrated in the spin-$\tfrac{1}{2}$ case that these difficulties can be circumvented by taking into account the distributional characteristics of the Poincar\'{e} charge operators and matrix elements, avoiding the necessity to define the wave-packet structure of the physical states themselves. These characteristics arise as a consequence of the fact that in local formulations of QFT fields are defined to be operator-valued distributions which satisfy a series of physically motivated axioms, including locality and relativistic covariance~\cite{Streater_Wightman64,Haag96,Bogolubov_Logunov_Oksak_Todorov90}. Since these axioms are assumed to hold independently of the coupling regime, this framework allows one to derive genuine non-perturbative constraints in a purely analytic manner. The main conclusion of~\cite{Lowdon_Chiu_Brodsky17} was that the zero momentum transfer limit of the leading two GFFs in the spin-$\tfrac{1}{2}$ EMT matrix element decomposition are completely constrained by the Poincar\'e transformation and on-shell properties of the states. This raises an important question: does this characteristic continue to hold for higher spin states, and if so, how is this limit affected by the spin of the states? The main goal of this work will be to address this question. As a by-product, by relating these leading GFFs to generalised parton distributions (GPDs), which can be in principle be accessed in processes such as deeply virtual Compton scattering (DVCS)~\cite{Ji:1998pc,Diehl:2003ny,Belitsky:2005qn}, the generalisation of the well-known spin-$\tfrac{1}{2}$ sum rules can be analysed for arbitrary spin states.\\

\noindent
The remainder of this paper is structured as follows: in Sec.~\ref{decomp} we define the leading terms which appear in the decomposition of the EMT matrix elements for massive states of arbitrary spin. Using this decomposition in Sec.~\ref{JK_decomp} we then apply the procedure developed in~\cite{Lowdon_Chiu_Brodsky17} to the angular momentum and boost matrix elements, and outline the subsequent constraints on the GFFs. In Sec.~\ref{EM_PL_decomp} we generalise this approach to the covariant Lorentz generators, and discuss the implications of these results in Sec.~\ref{appl}. Finally, in Sec.~\ref{concl} we conclude by summarising our key findings.

\newpage

\section{Gravitational form factors for arbitrary spin states}
\label{decomp}

In order to analyse the constraints imposed on the GFFs appearing in the decomposition of the EMT matrix elements for states of arbitrary spin, one must first outline how these states are defined. Due to the distributional nature of quantised fields it follows that any definite momentum eigenstate $|p \rangle$ is in fact a \textit{distributional}-valued state~\cite{Bogolubov_Logunov_Oksak_Todorov90}. Normalisable states $|g\rangle = \int \ud^{4}p \, g(p)|p \rangle$ are constructed by integrating $|p \rangle$ with test functions (or wave-packets) $g(p)$, chosen to belong to the space of Schwartz functions of fast decrease $\mathcal{S}(\mathbb{R}^{1,3})$. As will be discussed later, this choice of test functions also plays an important role in the definition of charge operators. For the purposes of the analysis in this paper we will be concerned only with massive physical on-shell states. Since $|p \rangle$ is \textit{a priori} defined for any four-momentum $p \in \mathbb{R}^{1,3}$, one can impose this requirement by considering eigenstates which are restricted to the upper hyperboloid $\Gamma_{M}^{+}=\{p \in \mathbb{R}^{1,3}:p^{2}=M^{2}, p^{0}>0 \}$ as follows   
\begin{align}
|p ,m;M \rangle = \delta_{M}^{(+)}(p)|p,m\rangle \equiv 2\pi \,\theta(p^{0})\,\delta(p^{2}-M^{2}) |p,m\rangle,
\end{align}
where $M$ is the mass of the state and $m$ is the canonical spin projection in the $z$-direction. As a result, even if the test function $g(p)$ has support outside of the mass shell, the normalisable state $|g\rangle$ satisfies the mass shell constraint. Since the norm of the unrestricted eigenstate $|p , m \rangle$ is given by $\langle p',m'|p ,m \rangle = 2p^{0}\,(2\pi)^{3}\delta^{3}(\uvec{p}'-\uvec{p})\,\delta_{m'm}$, the above definition implies that the inner product of the on-shell states has the following Lorentz-covariant form 
\begin{align}
\langle p',m';M|p ,m;M \rangle = (2\pi)^{4}\delta^{4}(p'-p)\,\delta_{M}^{(+)}(p)\,\delta_{m'm}.
\label{norm}
\end{align} 
Now it remains to parametrise the EMT matrix elements with respect to these states. Taking the EMT operator $T^{\mu\nu}$ to be symmetric it follows from the conservation of this current, together with the Lorentz covariance and discrete space-time symmetries, that the matrix elements for arbitrary spin can be written~\cite{Boulware:1974sr}
\begin{align}
\langle p',m';M|T^{\mu\nu}(0)|p ,m;M \rangle = \overline{\eta}_{m'}(p')O^{\mu\nu}(p',p)\eta_{m}(p)  \, \delta_{M}^{(+)}(p')\,\delta_{M}^{(+)}(p),
\label{T_decomp}
\end{align}
with the Lorentz covariant factor 
\begin{align}
O^{\mu\nu}(p',p)= \bar{p}^{\{\mu}\bar{p}^{\nu\}}  A(q^{2}) + i \bar{p}^{\{\mu}S^{\nu\}\rho}q_{\rho} \, G(q^{2}) + \cdots
\label{T_decomp_2}
\end{align}
The $\cdots$ indicates contributions with an explicitly higher-order dependence on the four-momentum transfer $q=p'-p$. We define the average four-momentum: $\bar{p}= \tfrac{1}{2}(p'+p)$ and the symmetrisation: $a^{\{\mu }b^{\nu\}}= a^{\mu}b^{\nu}+a^{\nu}b^{\mu}$. $S^{\mu\nu}$ are the Lorentz generators in the chosen spin representation and $\eta_{m}(p)$ are the arbitrary spin generalisation of the spinor and polarisation vector in the half-odd and half-even spin cases respectively. In particular, in the spin-$\tfrac{1}{2}$ case one has that: $S^{\mu\nu}= \frac{i}{4}\!\left[\gamma^{\mu},\gamma^{\nu}\right]$, $\eta_{m}(p)\propto u_{m}(p)$, and Eq.~\eqref{T_decomp} agrees\footnote{Here we have chosen to define a single form factor $G(q^{2})$ for the component involving the Lorentz generator, so $G(q^{2}) = A(q^{2})+B(q^{2})$ in comparison with~\cite{Lowdon_Chiu_Brodsky17} for the spin-$\tfrac{1}{2}$ case.} with the well-known matrix element parametrisation of the nucleon EMT~\cite{Ji:1996ek}. The parametrisation~\eqref{T_decomp} also assumes that the covariant density matrix $[\rho_{m'm}(p)]^A_{\phantom{A}B} = [\eta_{m}(p)]^A[\overline{\eta}_{m'}(p)]_B$ has the mass-independent normalisation
\begin{align}
\text{Tr}[\rho_{m'm}(p)] =[\eta_{m}(p)]^A[\overline{\eta}_{m'}(p)]_A= \overline{\eta}_{m'}(p)\eta_{m}(p)= \delta_{m'm}.
\end{align}
Note that the trace is performed in the spin representation space only. A characteristic feature of Eq.~\eqref{T_decomp} is that the arbitrary spin $\eta_{m}(p)$ appear in a purely external manner, and that the complexity of this expression is determined by the possible combinations of contracting $\bar{p}^{\mu}$ and $S^{\mu\nu}$ with $q^{\mu}$, whilst respecting the conservation and symmetry of $T^{\mu\nu}$. Eq.~\eqref{T_decomp} also makes it manifest that the non-perturbative structure is completely encoded in the GFFs. \\

\noindent
Now that the structure of the EMT matrix elements for arbitrary spin has been determined, in the proceeding sections we will apply an analogous approach to~\cite{Lowdon_Chiu_Brodsky17} in order to derive constraints on the GFFs $A(q^2)$ and $G(q^2)$. In~\cite{Boulware:1974sr} these constraints were outlined using a perturbative gravitational approach together with the Rarita-Schwinger representation. Two other derivations were also proposed in~\cite{Cho:1976de}: one based on the expansion of the EMT in momentum space\footnote{We observe that this first derivation contains a loophole since a possible contribution of the type $\bar p^\mu \bar p^\nu J^{0\rho}q_\rho$, allowed owing to Wigner rotation effects~\cite{Lorce:2017isp}, has not been considered. This is in line with~\cite{Bakker_Leader_Trueman04}, where the fallacy of the expansion used in~\cite{Cho:1976de} (and also later in~\cite{Shore:1999be}) was pointed out.}, and another using Schwinger's multispinor formalism together with a non-relativistic expansion. In contrast to these former works the proof we provide is purely non-perturbative, and relies only on the Poincar\'e invariance of the QFT and the distributional properties of the matrix elements. Moreover, this approach properly takes into account Wigner rotation effects, and does not require one to consider a non-relativistic expansion or a particular massive spin representation of the states.

\section{Lorentz generator matrix elements}
\label{JK_decomp}

In~\cite{Lowdon_Chiu_Brodsky17} it was first demonstrated that one can derive constraints on the GFFs by performing a distributional matching procedure. This procedure involves comparing the parametrisation of the matrix elements of the Poincar\'{e} charges with the representation that results from the explicit action of these charges on the states. Due to the distributional nature of the Poincar\'{e} currents, a rigorous definition of the corresponding charges requires integration with a sequence of appropriate test functions. As will be emphasised in the calculations that follow, taking into account the subtleties of these charge definitions is essential for obtaining consistent form factor constraints. A more detailed discussion of the motivation behind the various charge definitions can be found in~\cite{Lowdon_Chiu_Brodsky17} and references within.

\subsection{Angular momentum matrix element}

Let us start with the angular momentum operator $J^{i}$. Its rigorous definition reads
\begin{align}
J^{i} = \frac{1}{2}\,\epsilon^{ijk} \lim_{\substack{d \rightarrow 0 \\ R \rightarrow \infty}}\int \ud^{4}x \, f_{d,R}(x) \left[ x^{j}T^{0k}(x) - x^{k}T^{0j}(x) \right], \label{j_charge}
\end{align}
where $f_{d,R}(x) \equiv \alpha_{d}(x^{0})F_{R}(\uvec{x}) \in \mathcal{S}(\mathbb{R}^{1,3})$, and the test functions $\alpha_{d}$, $F_{R}$ satisfy the conditions
\begin{align}
\int \ud x^{0} \, \alpha_{d}(x^{0}) &=1, &\alpha_{d}(x^{0})& \xrightarrow{d \rightarrow 0} \delta(x^{0}), \label{test1} \\
F_{R}(\uvec 0) &= 1, &F_{R}(\uvec{x}) &\xrightarrow{R \rightarrow \infty} 1. \label{test2}
\end{align}
This definition guarantees that $J^{i}$ is convergent within matrix elements, and also independent of the specific choice of test functions used in the limit\footnote{The independence of the choice of temporal test function $\alpha_{d}$ can in particular be interpreted as the quantum generalisation of the time independence of the charge~\cite{Strocchi13}.}. Using this definition and a translation of the EMT operator $T^{\mu\nu}(x)=e^{iP\cdot x}\,T^{\mu\nu}(0)\,e^{-iP\cdot x}$, it follows that the angular momentum matrix element between the states $|p ,m ;M\rangle$ can be written
\begin{align}
\langle p',m';M|J^{i}|p ,m ;M\rangle &= \epsilon^{ijk} \lim_{\substack{d \rightarrow 0 \\ R \rightarrow \infty}} \int \ud^{4}x \, f_{d,R}(x)\, x^{j}e^{iq\cdot x} \,\langle p',m';M|T^{0k}(0)|p,m ;M\rangle \nonumber  \\
&= -i\epsilon^{ijk} \lim_{\substack{d \rightarrow 0 \\ R \rightarrow \infty}} \frac{\partial \widetilde{f}_{d,R}(q)}{\partial q_{j}}\, \langle p',m';M|T^{0k}(0)|p ,m ;M\rangle,
\label{j_eq}
\end{align}
with $\widetilde{f}_{d,R}(q)=\int \ud^{4}x \, e^{iq\cdot x} f_{d,R}(x)$. From the conditions in Eqs.~\eqref{test1} and~\eqref{test2} one has that
\begin{align}
\lim_{\substack{d \rightarrow 0 \\ R \rightarrow \infty}}\widetilde{f}_{d,R}(q) = (2\pi)^3\delta^3(\uvec{q}),
\label{dR_lim}
\end{align}  
which due to Eq.~\eqref{j_eq} implies that one must determine the product of derivatives of delta $\partial^j\delta^3(\uvec q)=\frac{\partial}{\partial q_j}\delta^3(\uvec q)$ and other factors in order to evaluate the full matrix element. The general form for this type of covariant distributional expression is derived in Appendix~\ref{appendix_a}. In particular, using the parametrisation in Eq.~\eqref{T_decomp} together with Eq.~\eqref{cov_dist} it follows that Eq.~\eqref{j_eq} can be written
\begin{align}
\langle p',m';M|J^{i}|p ,m ;M\rangle = (2\pi)^{4}\delta_{M}^{(+)}(\bar p)\, \mathcal J^i_{m'm}(\bar p,q),
\label{Jampl}
\end{align}
with the reduced matrix element
\begin{align}
\mathcal{J}^{i}_{m'm}(\bar{p},q) &= -i\epsilon^{ijk}\bar{p}^{k} \left[ \delta_{m'm} \, \partial^{j}\delta^4(q) - \left.\partial^{j} \!\left[\overline{\eta}_{m'}(p')\eta_{m}(p)\right]\right|_{q=0} \delta^4(q)\right]  A(q^2) \nonumber \\
&\quad +\frac{1}{2}\,\epsilon^{ijk} \left[ \overline{\eta}_{m'}(\bar{p})S^{jk}\eta_{m}(\bar{p}) \right] \delta^4(q) \,    G(q^2).
\label{rel_i}
\end{align}
It is important to note that the temporal derivative of $\delta^{4}(q)$ which can potentially appear due to Eq.~\eqref{cov_dist} drops out of this expression due to the contraction with $\epsilon^{ijk}\bar{p}^{k}$. In order to further simplify this expression one needs to evaluate the derivative term for arbitrary spin. As proved in Appendix~\ref{appendix_b}, it turns out that one has the following closed form expression
\begin{align}
\left.\frac{\partial}{\partial q_{j}}\left[\overline{\eta}_{m'}(p')\eta_{m}(p)\right]\right|_{q=0} = \frac{i}{|\bar{\uvec{p}}|^{2}}\,\epsilon^{jlr}\bar{p}^{l}\left[\Sigma^r_{m'm}(\bar p)-\Sigma^r_{m'm}(k)\right], \label{rel}
\end{align}
where $k^{\mu} = Mg^{\mu 0}$ is rest-frame four-momentum, and $\Sigma^i_{m'm}(\bar p)=\text{Tr}\left[\rho_{m'm}(\bar p)\Sigma^{i}\right]$ with $\Sigma^i=\frac{1}{2}\epsilon^{ijk}S^{jk}$ the spin matrices in the chosen spin representation. Inserting Eq.~\eqref{rel} into Eq.~\eqref{rel_i} then gives
\begin{align}
\mathcal{J}^{i}_{m'm}(\bar{p},q) &=  \left[\Sigma^i_{m'm}(k) -\delta_{m'm}\,i \epsilon^{ijk}\bar{p}^{k}\partial^{j}\right]\delta^4(q)\,  A(q^{2}) \nonumber \\
& \quad + \Sigma^{i}_{m'm}(\bar{p})\, \delta^4(q)\left[  G(q^{2}) - A(q^{2}) \right],
\label{J_final_eq}
\end{align}
where we used the fact that helicity (i.e. spin projection along momentum) is invariant under longitudinal boosts: $\hat{\bar{\uvec p}}\cdot\uvec\Sigma_{m'm}(\bar p)=\hat{\bar{\uvec p}}\cdot\uvec\Sigma_{m'm}(k)$, with $\hat{\bar{\uvec p}}=\bar{\uvec p}/|\bar{\uvec p}|$. \\

\noindent
To derive constraints on $A(q^{2})$ and $G(q^{2})$ one can observe that due to the transformation properties of the states $|p,m;M\rangle$ under rotations, the $J^{i}$ reduced matrix elements must have the general form
\begin{align}
\mathcal J^i_{m'm}(\bar p,q) = \left[\Sigma^i_{m'm}(k) -\delta_{m'm} \,i\epsilon^{ijk}\bar{p}^{k}\partial^{j} \right]\delta^4(q),
\label{j_eq_rot}
\end{align}
which is derived in Appendix~\ref{appendix_a}. Since Eqs.~\eqref{J_final_eq} and~\eqref{j_eq_rot} are simply different representations of the same matrix element, the coefficients of these distributions must coincide, which requires that the following identities hold  
\begin{align}
A(q^{2})\,\delta^4(q) &= \delta^4(q), \label{constr1} \\
A(q^{2})\,\partial^{j}\delta^4(q) &= \partial^{j}\delta^4(q), \label{constr2} \\
\left[  G(q^{2}) -  A(q^{2}) \right] \delta^4(q) &= 0. \label{constr3}
\end{align}
Combining these identities implies the constraint\footnote{Technically $A(q^{2})$ and $G(q^{2})$ are distributions in $q$, and so are in general not point-wise defined. Nevertheless, one can interpret $A(0)$ and $G(0)$ using a limiting procedure~\cite{Lowdon_Chiu_Brodsky17}.} 
\begin{align}
A(0) = G(0) = 1,
\label{FF_equality}
\end{align}
which proves that the $q \rightarrow 0$ behaviour of the two leading GFFs in the EMT matrix element decomposition is completely independent of both the spin and internal composition of the states, and that the actual limiting values of these form factors coincide.

\subsection{Boost matrix element}

Another important point which was raised in~\cite{Lowdon_Chiu_Brodsky17} is that the constraints on the GFFs are not specifically related to the conservation of angular momentum, contrary to what is often thought in the literature. To emphasise this point it was demonstrated (for spin $\tfrac{1}{2}$) that identical form factor constraints can also be obtained using the matrix elements of the boost generators $K^{i}$. Since the calculations in the preceding section concluded that $A(0) = G(0) = 1$ is a spin-independent constraint, one would therefore expect that the same constraint must also arise from the structure of $\langle p',m';M|K^{i}|p ,m ;M\rangle$. It turns out that this is in fact the case, as will be demonstrated in the remainder of this section. \\

\noindent
Similarly to $J^{i}$ the boost generator is rigorously defined by
\begin{align}
K^{i} =  \lim_{\substack{d \rightarrow 0 \\ R \rightarrow \infty}}\int \ud^{4}x \, f_{d,R}(x) \left[ x^{0}T^{0i}(x) - x^{i}T^{00}(x) \right], 
\label{k_charge} 
\end{align}
and hence the boost matrix element can be written in the form
\begin{align}
\langle p',m';M|K^{i}|p ,m ;M\rangle = i \lim_{\substack{d \rightarrow 0 \\ R \rightarrow \infty}} \frac{\partial \widetilde{f}_{d,R}(q)}{\partial q_{i}}\, \langle p';m';M|T^{00}(0)|p ;m ;M\rangle.
\label{k_eq}
\end{align} 
The term proportional to $\langle p',m';M|T^{0i}(0)|p ,m ;M\rangle$ vanishes due to the definition of the test functions in Eqs.~\eqref{test1} and~\eqref{test2}
\begin{align}
\lim_{\substack{d \rightarrow 0 \\ R \rightarrow \infty}} \int \ud^{4}x \, e^{iq \cdot x}x^{0} \, \alpha_{d}(x^{0}) F_R(\uvec x) = 0.  
\end{align}
Inserting the parametrisation~\eqref{T_decomp} and using Eq.~\eqref{cov_dist} one can write
\begin{align}
\langle p',m';M|K^{i}|p ,m ;M\rangle = (2\pi)^{4}\delta_{M}^{(+)}(\bar p)\, \mathcal K^i_{m'm}(\bar p,q),
\label{Kampl}
\end{align}
with the reduced matrix element
\begin{align}
\mathcal{K}^{i}_{m'm}(\bar{p},q) &= i \left[\delta_{m'm}\,(\bar{p}^{0} \partial^{i} - \bar{p}^{i} \partial^{0})\delta^4(q) - \left.\bar{p}^{0}\partial^{i}\!\left[\overline{\eta}_{m'}(p')\eta_{m}(p)\right]\right|_{q=0}\delta^4(q)       \right]  A(q^{2}) \nonumber \\
& \quad +\left[\overline{\eta}_{m'}(\bar{p})S^{0i}\eta_{m}(\bar{p})  \right] \delta^4(q)  \, G(q^{2}).
\label{boost_eval}
\end{align}  
Due to Eq.~\eqref{rel} one sees that in order to further simplify this relation one requires an explicit expression for $\epsilon^{ijk}\bar{p}^{j}\bar{p}^{0} \, \Sigma^k_{m'm}(\bar p)$. As shown in Appendix~\ref{appendix_b}, due to the properties of the covariant density matrix $\rho_{m'm}(\bar{p})$ one can prove that  
\begin{align}
\epsilon^{ijk}\bar{p}^{j}\bar{p}^{0} \, \Sigma^k_{m'm}(\bar{p})  = -|\bar{\uvec{p}}|^{2} \, \kappa^i_{m'm}(\bar{p})+ M \epsilon^{ijk}\bar{p}^{j} \, \Sigma^k_{m'm}(k), 
\label{J_contract}
\end{align} 
where $\kappa^i_{m'm}(\bar{p})=\text{Tr}\left[\rho_{m'm}(\bar p)\kappa^{i}\right]$ with $\kappa^i=S^{0i}$ the boost generator matrices in the chosen spin representation. Upon insertion into Eq.~\eqref{boost_eval} this finally gives 
\begin{align}
\mathcal{K}^{i}_{m'm}(\bar{p},q) &=   \left[-\frac{\epsilon^{ijk}\bar{p}^{j}}{\bar{p}^{0}+M}\,\Sigma_{m'm}^{k}(k)  +\delta_{m'm}\, i(\bar{p}^{0} \partial^{i} - \bar{p}^{i} \partial^{0})\right]\delta^4(q)\, A(q^{2}) \nonumber \\
& \quad + \kappa^i_{m'm}(\bar{p})\,\delta^4(q) \left[G(q^{2})- A(q^{2})\right].
\label{K_final_eq} 
\end{align}
Just as the rotation transformation properties of the states $|p,m;M\rangle$ were used to constrain the matrix elements of $J^{i}$, one can perform an analogous procedure for boosts. In this case the $K^{i}$ reduced matrix elements have the general form  
\begin{align}
\mathcal{K}^{i}_{m'm}(\bar{p},q) = \left[-\frac{\epsilon^{ijk}\bar{p}^{j}}{\bar{p}^{0}+M}\,\Sigma_{m'm}^{k}(k)   +\delta_{m'm}\, i(\bar{p}^{0} \partial^{i} - \bar{p}^{i} \partial^{0})\right]\delta^4(q),
\label{m_eq_boost}
\end{align}
which is derived in Appendix~\ref{appendix_a}. Comparing this with Eq.~\eqref{K_final_eq} one immediately sees that the equality of these expressions implies the same relations\footnote{Due to the non-vanishing $\partial^{0}\delta^{4}(q)$ term one also has the constraint: $A(q^{2})\,\partial^{0}\delta^{4}(q) =\partial^{0}\delta^{4}(q)$. However, this relation is essentially trivial because it implies $\partial^{0}A(0)=0$, which follows immediately from the fact that $A(q^{2})$ depends only on $q^{2}$.} as in Eqs.~\eqref{constr1},~\eqref{constr2} and~\eqref{constr3}. As anticipated, this result emphasises that the form factor constraints are not specific to the properties of any one Lorentz generator. In the next section, we will instead consider the constraints imposed on $A(q^{2})$ and $G(q^{2})$ by the \textit{covariant} Poincar\'e generators: the four-momentum operator $P^{\mu}$, and the covariant generalisations of $J^{i}$ and $K^{i}$.

\section{Covariant generator matrix elements}
\label{EM_PL_decomp}

Before discussing the covariant generalisation of the rotation and boost operators, consider the simplest case of the four-momentum operator $P^{\mu}$. Although $P^{\mu}$ does not involve an explicit factor of $x^{\alpha}$ in its definition, $P^{\mu}$ is nevertheless defined by smearing with the same class of test functions as the Lorentz generators
\begin{align}
P^{\mu} =  \lim_{\substack{d \rightarrow 0 \\ R \rightarrow \infty}}\int \ud^{4}x \ f_{d,R}(x)  \, T^{0\mu}(x)  \label{mom_charge}
\end{align}
and hence the $P^{\mu}$ matrix element can be written as
\begin{align}
\langle p, m';M|P^{\mu}|p, m ;M\rangle = \lim_{\substack{d \rightarrow 0 \\ R \rightarrow \infty}} \widetilde{f}_{d,R}(q) \, \langle p', m';M|T^{0 \mu}(0)|p, m ;M\rangle.
\label{p_eq}
\end{align} 
Inserting the form factor decomposition of Eq.~(\ref{T_decomp}) and applying Eq.~\eqref{cov_dist_simple} it immediately follows that
\begin{align}
\langle \bar{p}+ \tfrac{1}{2}q,m';M| P^{\mu}|\bar{p}- \tfrac{1}{2}q,m;M\rangle = (2\pi)^{4} \delta_{M}^{(+)}(\bar{p}) \, \bar{p}^{\mu} \, \delta_{m'm} \, A(q^{2})  \delta^{4}(q), 
\end{align}
where the $G(q^{2})$-dependent terms have dropped out due to the explicit $q$ factor. Since the on-shell states are defined to have an inner product as in Eq.~(\ref{norm}), and $|p, m;M \rangle$ are momentum eigenstates satisfying: $P^{\mu}|p, m; M \rangle = p^{\mu}|p, m; M \rangle$, these relations therefore imply the spin-independent constraint
\begin{align}
A(q^{2})\, \delta^{4}(q) = \delta^{4}(q),
\end{align}
which is simply $A(0)=1$. 

\subsection{Pauli-Lubanski matrix element}
 
The covariant generalisation of $J^{i}$, the Pauli-Lubanski operator $W^{\mu}$, is defined by\footnote{Here we use the convention $\epsilon_{0123}=+1$.} 
\begin{align}
W^{\mu} &= \frac{1}{2}\epsilon^{\mu}_{\phantom{\mu} \rho\sigma\lambda}M^{\rho\sigma}P^{\lambda}. 
\end{align}
By definition, the rest-frame matrix element of $W^{\mu}$ coincides with $J^{i}$, up to an overall mass coefficient. Before calculating the matrix element of $W^{\mu}$ one must first define the general Lorentz generator $M^{\mu\nu}$. Similarly to $J^{i}$ and $K^{i}$ one has that
\begin{align}
M^{\mu\nu} =  \lim_{\substack{d \rightarrow 0 \\ R \rightarrow \infty}}\int \ud^{4}x \, f_{d,R}(x) \left[ x^{\mu}T^{0\nu}(x) - x^{\nu}T^{0\mu}(x) \right]. 
\label{cov_charge}
\end{align}
The matrix element of $W^{\mu}$ can then be written
\begin{align}
\langle p',m';M|W^{\mu}|p ,m ;M\rangle = -i\epsilon^{\mu}_{\phantom{ \mu} \rho\sigma\lambda} \, p^{\lambda}  \lim_{\substack{d \rightarrow 0 \\ R \rightarrow \infty}} \frac{\partial \widetilde{f}_{d,R}(q)}{\partial q_{\rho}}\, \langle p',m';M|T^{0\sigma}(0)|p ,m ;M\rangle,
\end{align}
which after inserting the parametrisation~\eqref{T_decomp} and applying Eq.~\eqref{cov_dist} gives
\begin{align}
\langle p',m';M|W^{\mu}|p ,m ;M\rangle = (2\pi)^{4}\delta_{M}^{(+)}(\bar p)\, \mathcal W^\mu_{m'm}(\bar p,q),
\label{Wampl}
\end{align}
with the reduced matrix element
\begin{align}
\mathcal W^\mu_{m'm}(\bar p,q)  =  S^\mu_{m'm}(\bar p)\, \delta^4(q) \, G(q^{2}), \label{g_eq}
\end{align}
where $S^\mu_{m'm}(\bar{p})=\text{Tr}\left[\rho_{m'm}(\bar p)S^\mu\right]$ with $S^{\mu}=\frac{1}{2}\epsilon^\mu_{\phantom{\mu}\rho\sigma\lambda}S^{\rho\sigma}\bar p^\lambda$ the covariant spin matrices\footnote{The explicit form for the covariant spin matrices in terms of the non-conserved ones $\Sigma_{m'm}^{i}(k)$ is given by: $S^{\mu}_{m'm}(p) = \left( \uvec{p}\cdot  \boldsymbol{\Sigma}_{m'm}(k), \, M \boldsymbol{\Sigma}_{m'm}(k) +  \frac{ \uvec{p}\cdot \boldsymbol{\Sigma}_{m'm}(k)}{p^{0}+M}\uvec{p}  \right)$.} in the chosen spin representation. The dependence on $A(q^{2})$ completely drops out due to the contraction with $\epsilon^{\mu}_{\phantom{\mu} \rho\sigma\lambda} \, \bar p^{\lambda}$ and the explicit $\bar{p}^{\sigma}$ factor multiplying this term. Unlike the rotation and boost generators the Pauli-Lubanski operator acts in a diagonal manner on the momentum component of the states, and so the reduced matrix elements have the general form
\begin{align}
\mathcal W^\mu_{m'm}(\bar p,q)  = S^{\mu}_{m'm}(\bar p)\,\delta^4(q).
\label{w_eq}
\end{align}  
Equating Eqs.~\eqref{g_eq} and~\eqref{w_eq} immediately implies the constraint  
\begin{align}
G(q^{2})\,\delta^4(q) = \delta^4(q),
\end{align} 
which is nothing more than the condition $G(0)=1$. 

\subsection{Covariant boost matrix element}

The covariant boost $B^{\mu}$ is defined by the symmetrised expression
\begin{align}
B^{\mu} = \frac{1}{2}\left[ M^{\nu\mu}P_{\nu} + P_{\nu}M^{\nu\mu} \right],
\end{align}
and coincides with $K^{i}$ within matrix elements of rest-frame states. The general matrix elements of $B^{\mu}$ can be directly related to those of the rotation and boost operators\footnote{Given these definitions of $W^{\mu}$ and $B^{\mu}$ the general Lorentz generator can be written in the following form: $M^{\mu\nu} = -\tfrac{1}{2P^{2}}\left[\{B^{\mu},P^{\nu}\} - \{B^{\nu},P^{\mu}\}\right] -\tfrac{1}{P^{2}}\epsilon^{\mu\nu}_{\phantom{\mu}\phantom{\mu} \alpha \beta} W^{\alpha}P^{\beta}$, where $\{\cdot,\cdot\}$ is the anti-commutator.}, and in particular the corresponding reduced matrix elements $\mathcal{B}^{\mu}_{m'm}(\bar{p},q)$ are given by
\begin{align}
\mathcal{B}^{0}_{m'm}(\bar{p},q) &= \bar{p}^{i} \, \mathcal{K}^{i}_{m'm}(\bar{p},q), \label{reducedB1} \\
\mathcal{B}^{i}_{m'm}(\bar{p},q) &= \bar{p}^{0} \, \mathcal{K}^{i}_{m'm}(\bar{p},q) +\epsilon^{ijk}\bar{p}^{j} \, \mathcal{J}^{k}_{m'm}(\bar{p},q).
\label{reducedB2}
\end{align}
Using Eq.~\eqref{K_final_eq} it follows that Eq.~\eqref{reducedB1} can be written 
\begin{align}
\mathcal{B}^{0}_{m'm}(\bar{p},q) &=  
i \delta_{m'm}\, \bar{p}^{i} \left[\bar{p}^{0}\partial^{i}  - \bar{p}^{i} \partial^{0}\right]\delta^{4}(q)\, A(q^{2}) + \bar{p}^{i}\kappa^i_{m'm}(\bar{p})\,\delta^{4}(q) \left[G(q^{2})- A(q^{2})\right] \nonumber \\
&=  i\delta_{m'm}\, \bar{p}^{i}\left[\bar{p}^{0}\partial^{i}   - \bar{p}^{i} \partial^{0}\right]\delta^{4}(q)\, A(q^{2}), 
\end{align}
where the last line follows from the fact that: $\bar{p}^{i}\kappa^{i}_{m'm}(\bar{p})= \bar{p}^{i}\kappa^{i}_{m'm}(k)= 0$. Comparing this with the general boost matrix element representation in Eq.~\eqref{m_eq_boost} projected on $\bar{p}$ therefore implies the constraints
\begin{align}
A(q^{2})\,\partial^{j}\delta^4(q)  = \partial^{j}\delta^4(q), \hspace{5mm} A(q^{2})\,\partial^{0}\delta^4(q) = \partial^{0}\delta^4(q). \label{constrB}
\end{align}
Similarly, computing $\mathcal{B}^{i}_{m'm}(\bar{p},q)$ one obtains  
\begin{align}
\mathcal{B}^{i}_{m'm}(\bar{p},q) &=  \frac{M\epsilon^{ijk}\bar{p}^{j}}{\bar{p}^{0}+M}\,\Sigma_{m'm}^{k}(k) \, \delta^{4}(q)\, A(q^{2}) + \delta_{m'm}\, i \left[  \bar{p}^{2}\partial^{i} -\bar{p}^{i}(\bar{p} \cdot \partial)\right]\delta^4(q)\, A(q^{2}) \nonumber \\
& \quad + \left[ \epsilon^{ijk}\bar{p}^{j}\Sigma_{m'm}^{k}(\bar{p}) +\bar{p}^{0}\kappa^i_{m'm}(\bar{p}) \right] \delta^{4}(q) \left[G(q^{2})- A(q^{2})\right] \nonumber \\
&=   \frac{M\epsilon^{ijk}\bar{p}^{j}}{\bar{p}^{0}+M}\,\Sigma_{m'm}^{k}(k) \, \delta^{4}(q)\, A(q^{2}) + \delta_{m'm}\, i \left[  \bar{p}^{2}\partial^{i} -\bar{p}^{i}(\bar{p} \cdot \partial)\right]\delta^4(q)\, A(q^{2}),
\end{align}
where the last line follows from Eq.~\eqref{id1} derived in Appendix~\ref{appendix_b}. Comparing this with both the general Lorentz generator matrix elements in Eqs.~\eqref{j_eq_rot} and~\eqref{m_eq_boost} one is left with the constraints in Eq.~\eqref{constrB}, together with the condition: $A(q^{2})\,\delta^{4}(q) = \delta^{4}(q)$. \\

\noindent
The calculations in this section explicitly demonstrate that the matrix elements of the covariantised rotation and boost operators \textit{separately} determine the constraints on $G(q^{2})$ and $A(q^{2})$ respectively. In other words, choosing this covariant operator basis results in a diagonalisation of the constraints. Overall, it initially appears that the matrix elements of the Lorentz generators, or their covariantised versions, are sufficient to derive all of the form factor constraints. Since these constraints follow from the Lorentz transformations properties of the states, this seemingly suggests that only Lorentz symmetry is involved. However, in deriving the matrix element equations we have also implicitly used the spacetime translation transformation: $e^{iP\cdot x}|p,m;M\rangle = e^{ip\cdot x}|p,m;M\rangle$. This explains why the condition $A(q^{2})\,\delta^{4}(q) = \delta^{4}(q)$, which follows from the matrix element of $P^{\mu}$, is also implied by the matrix elements of the various Lorentz generators. Ultimately this means that the total constraints on the GFFs are a result of the \textit{full} Poincar\'e symmetry, together with the on-shell restriction of the states.

\section{Applications} 
\label{appl}

We now turn to the phenomenological implications of our results, focussing specifically on the applications to hadronic physics. The quantum interactions between matter and gravity are in principle encoded in the GFFs, but in practice they are too weak to be directly measured in experiment. One way of accessing information about QCD matter is through the generalised parton distributions (GPDs)~\cite{Ji:1996ek,Ji:1998pc,Diehl:2003ny,Belitsky:2005qn}. In this case one is dealing with a non-local operator along the light-like direction $n$, which enters into the description of deeply virtual Compton scattering (DVCS) at the amplitude level. The leading-twist quark and gluon GPDs have the following form:
\begin{align}
&V_{m'm}^{ q} = \frac{1}{2} \int_{-\infty}^\infty \frac{\ud{z}}{2\pi} e^{ix(\bar{p}\cdot n)z} \left\langle p', m'; M \middle| \overline{\psi}\!\left(-\tfrac{z}{2}n\right) (\gamma\cdot n) \mathcal{W}_{\left[-\frac{z}{2}n,\frac{z}{2}n\right]} \psi\!\left(\tfrac{z}{2}n\right) \middle| p, m; M \right\rangle,    \label{GPDcorrelators} \\
&V_{m'm}^{g}  =   \frac{n_{\alpha}n_{\beta}}{2x(\bar{p}\cdot n)} \int_{-\infty}^\infty \frac{\ud{z}}{2\pi} e^{ix(\bar{p}\cdot n)z} \left\langle p', m'; M \middle|F_{\alpha}^{ \ \lambda} \!\left(-\tfrac{z}{2}n\right)\mathcal W_{ \left[-\frac{z}{2}n,\frac{z}{2}n\right]} F_{\lambda\beta}\!\left(\tfrac{z}{2}n\right) \middle| p, m; M \right\rangle,
\label{GPDcorrelatorq}
\end{align}
where $x$ is the longitudinal momentum fraction of the parton, and $\mathcal W_{[a,b]}$ denotes a straight Wilson line in the adjoint representation joining the spacetime points $a$ and $b$. The non-local quark and gluon operators which appear within the matrix elements of these definitions ($\mathcal{O}_{V}^{q}$ and $\mathcal{O}_{V}^{g}$) are related to the quark and gluon EMT operators via the second Mellin moment 
\begin{align}
\int_{-1}^1 \ud{x}\,  x\, \mathcal{O}_{V}^{q}  &=  \frac{ 1 }{4(\bar{p}\cdot n)^{2}}  \overline{\psi}(0) (\gamma\cdot n) (i\overset{\leftrightarrow}{D}\!\!\!\!\phantom{D}\cdot n) \psi(0)  =\frac{n_{\mu}n_{\nu}T_q^{\mu \nu}}{2(\bar{p}\cdot n)^2}, \label{eqn:gpd:moments} \\
\int_{-1}^1 \ud{x}\,x\, \mathcal{O}_{V}^{g}  &=  \frac{n_{\mu}n_{\nu}}{2(\bar{p}\cdot n)^2}  F^{\mu\lambda}(0) F_{\lambda}^{\ \nu}(0)  =  \frac{n_{\mu}n_{\nu}T_g^{\mu\nu}}{2(\bar{p}\cdot n)^2}, \label{operatorRelation}
\end{align}
with $\overset{\leftrightarrow}{D}_{\mu}= \overset{\rightarrow}{D}_{\mu}-\overset{\leftarrow}{D}_{\mu}$. Eq.~\eqref{T_decomp} is the most general decomposition for the symmetric and conserved EMT, and the terms we are interested in depend at most linearly on $q$. To spell out the relation between the GFFs in Eq.~\eqref{T_decomp} and the Mellin moments of GPDs we need to restrict ourselves to the twist-2 part of Eqs.~\eqref{eqn:gpd:moments} and~\eqref{operatorRelation}. In particular, neglecting terms with a higher power of $q$, the total GPD correlator at twist-2 reads
\begin{align}
\langle p',m';M|\mathcal{O}_{V}|p, m;M \rangle &= \overline{\eta}_{m'}(p') \!\left[ H_1(x,\xi,t) + \frac{iS^{\alpha\rho}n_{\alpha}q_{\rho}}{\bar{p}\cdot n}   \,H_2(x,\xi,t) + \cdots \right]\! \eta_{m}(p)  \, \delta_{M}^{(+)}(p)\delta_{M}^{(+)}(p'),
\label{GPD_decomp}
\end{align}
where $\xi=-(q \cdot n)/(2 \bar{p}\cdot n)$ is the light-front longitudinal momentum transfer, $t=q^2$, and $\mathcal{O}_{V}= \mathcal{O}_{V}^{q}+\mathcal{O}_{V}^{g}$. It follows that one can write the spin-independent relations
\begin{align}
\int_{-1}^1 \ud{x}\, x H_1(x,\xi,q^2)&=A(q^2) + \cdots , \label{GPD_rel1} \\
\int_{-1}^1 \ud{x}\, x H_2(x,\xi,q^2)&=G(q^2) + \cdots , \label{GPD_rel2}
\end{align}
where $\cdots$ denotes possible contributions arising from non-leading GFFs which are multiplied at least by $\xi^{2}$. Using the results derived in this paper one can now generalise Ji's sum rule~\cite{Ji:1996ek} such that it holds \textit{independently} of both the spin and structure of the hadron states. From Eq.~\eqref{FF_equality} it follows that for a state of arbitrary spin with longitudinal polarisation along, say, the $z$-direction, the total longitudinal linear and angular momentum (summed over quarks and gluons) reads:
\begin{align}
P^z &= \sum_{a=q,g}\int_{-1}^1 \ud{x}\, x H^a_1(x,0,0)= A(0)=1, \\
J^z &= \sum_{a=q,g}\int_{-1}^1 \ud{x}\, x H^a_2(x,0,0)= G(0)=1.
\label{Ji_sum}
\end{align}
\ \\ 
\noindent 
The totality of the structures that parametrise the EMT cannot be constrained by the action of the Poincar\'e generators alone, and in general contains both asymmetric and non-conserved terms. These terms are crucial in the study of the mechanical properties of hadrons, and receive different contributions from quarks and gluons. In particular, a general expression for Ji's relation which is valid for quarks and gluons separately would require the inclusion of such additional terms, as observed in~\cite{Cosyn:2019aio,Polyakov:2019lbq}. One approach to derive these terms is to write a parametrisation of the EMT for arbitrary spin states as an expansion in terms of spin multipoles\footnote{See~\cite{Cosyn:2019aio} for a discussion of the spin-1 case, and~\cite{Lorce:2009bs} for a parametrisation of the vector current case for arbitrary spin.}. \\

\noindent
Besides the hadronic relevance of the form factor constraints derived in this work, one can also interpret these conditions in a gravitational context. In particular, if one considers the situation in which the states correspond to a particle moving in an external (classical) gravitational field, the zero momentum transfer limit of the form factor $B(q^{2})= G(q^{2})-A(q^{2})$ has been argued to correspond to the anomalous gravitomagnetic moment (AGM) of the particle~\cite{Teryaev:1999su}, by analogy to the case of the anomalous magnetic moment of a charged particle. Due to the constraint in Eq.~\eqref{FF_equality} it follows immediately that $B(0)=0$, and hence with this interpretation the AGM must vanish for massive particles of \textit{any} spin. However, as previously outlined, this constraint arises purely from the Poincar\'e invariance of the theory, and does not in fact rely on any knowledge of the external gravitational interactions\footnote{Although the conditon $B(0)=0$ for arbitrary spin states has been discussed before~\cite{Boulware:1974sr,Cho:1976de,Teryaev:1999su}, until now this statement has not been proven in a non-perturbative manner.}. Einstein's equivalence principle is therefore not necessary to derive the constraint $B(0)=0$.

\section{Conclusions}
\label{concl}

The purpose of this work was to establish the most general constraints imposed on the form factors appearing in the Lorentz decomposition of the energy-momentum tensor matrix elements for massive states with arbitrary spin. By comparing the form factor representation of the angular momentum matrix elements with the representation due to the transformation properties of the states under rotations, we were able to prove that the $q \rightarrow 0$ behaviour of the leading two form factors $A(q^{2})$ and $G(q^{2})$ is completely independent of both the spin and internal structure of the states, and in particular that: $A(0)=G(0)=1$. Adopting an analogous procedure for the matrix elements of the boost generators $K^{i}$, we also established that the structure of these objects implies identical constraints to those derived using $J^{i}$. Together, these results emphasise that the constraints imposed on the leading gravitational form factors are not specifically related to the properties of any one of the Lorentz generators. Besides the standard Lorentz generators one can also use the covariantised version of these operators, the Pauli-Lubanski $W^{\mu}$ and covariant boost generator $B^{\mu}$, to derive constraints in the same manner. It turns out that $B^{\mu}$ and $W^{\mu}$ separately imply $A(0)=1$ and $G(0)=1$ respectively. In other words, choosing this covariant operator basis results in a diagonalisation of the constraints. The main conclusion from this analysis is that the spin-independent constraints on $A(q^{2})$ and $G(q^{2})$ are non-perturbative, and arise purely due to the general Poincar\'e transformation and on-shell properties of the states. These results have several immediate implications, including the spin-universality of Ji's sum rule for generalised parton distributions, and the vanishing of the anomalous gravitomagnetic moment for particles of any spin.

\section*{Acknowledgements}

This work was supported by the Agence Nationale de la Recherche under the projects No. ANR-18-ERC1-0002 and ANR-16-CE31-0019.

\appendix

\section{On-shell matrix elements of the Poincar\'{e} generators}
\label{appendix_a}

\subsection{Covariant representation}

The simplest on-shell matrix elements occur when calculating the matrix elements of the four-momentum operator. In this case one has distributional relations of the following form 
\begin{align}
\mathcal{T}(p',p) = \delta_{M}^{(+)}(p')\delta_{M}^{(+)}(p)\, C(p',p) \, \delta^{3}(\uvec{p}' -\uvec{p}),
\label{cov_dist_gen}
\end{align} 
where $C(p',p)$ is some function. In particular, $C(p',p)$ corresponds to the coefficients multiplying the form factors in Eq.~\eqref{T_decomp}. As with any distribution, the key to simplifying Eq.~\eqref{cov_dist_gen} is to understand how it acts on a generic test function $f$. For the purposes of the form factor analysis in this paper we are mainly interested in working with the variables $\bar{p}=\tfrac{1}{2}(p'+p)$ and $q=p'-p$. In these variables one can write the smeared distribution $\overline{\mathcal{T}}(\bar{p},q) = \mathcal{T}(p',p)$ in the following manner   
\begin{align}
\int \ud^{4}\bar{p} \,  \ud^{4}q \ \overline{\mathcal{T}}(\bar{p},q) \, f(\bar{p},q) &= \int \ud^{4}\bar{p} \,  \ud^{4}q \ \delta_{M}^{(+)}(\bar{p} + \tfrac{1}{2}q)\delta_{M}^{(+)}(\bar{p} - \tfrac{1}{2}q)\, \overline{C}(\bar{p},q) \, \delta^{3}(\uvec{q}) \, f(\bar{p},q) \nonumber \\ 
&= \int \ud^{4}\bar{p} \,  \ud^{4}q \ \frac{\delta \! \left(\bar{p}^{0}- \bar{p}^{0}_{\star} \right)\delta \! \left(q^{0}-q^{0}_{\star} \right)   }{4(\bar{p}^{0} +\tfrac{1}{2}q^{0})(\bar{p}^{0} -\tfrac{1}{2}q^{0}) } \overline{C}(\bar{p},q) \, \delta^{3}(\uvec{q}) \, f(\bar{p},q) \nonumber \\
&=  \int \ud^{3}\bar{\uvec{p}} \  \left.  \left( \frac{\overline{C}(\bar{p}^{0}_{\star}, \, \bar{\uvec{p}}, \, q^{0}_{\star}, \, \uvec{q}) \,  f(\bar{p}^{0}_{\star}, \, \bar{\uvec{p}}, \, q^{0}_{\star}, \, \uvec{q})}{4\sqrt{(\bar{\uvec{p}} + \tfrac{1}{2}\uvec{q})^2 +M^{2}}\sqrt{(\bar{\uvec{p}} - \tfrac{1}{2}\uvec{q})^2 +M^{2}}} \right) \right|_{\uvec{q}=\uvec{0}} \nonumber \\
&=  \int \ud^{3}\bar{\uvec{p}} \ \frac{\overline{C}(E_{\bar{\uvec{p}}}, \, \bar{\uvec{p}}, \, 0, \, \uvec{0}) \,  f(E_{\bar{\uvec{p}}}, \, \bar{\uvec{p}}, \, 0, \, \uvec{0})}{(2E_{\bar{\uvec{p}}})^{2}},
\label{P_deriv} 
\end{align}  
where $\overline{C}(\bar{p},q)=C(p',p)$ and one has used that
\begin{align}
\bar{p}^{0}_{\star} &= \frac{1}{2}\left[\sqrt{(\bar{\uvec{p}} + \tfrac{1}{2}\uvec{q})^2 +M^{2}} + \sqrt{(\bar{\uvec{p}} - \tfrac{1}{2}\uvec{q})^2 +M^{2}}\right], \label{eq_p0}\\
q^{0}_{\star} &= \sqrt{(\bar{\uvec{p}} + \tfrac{1}{2}\uvec{q})^2 +M^{2}} - \sqrt{(\bar{\uvec{p}} - \tfrac{1}{2}\uvec{q})^2 +M^{2}},  \label{eq_q0}
\end{align} 
which implies: $\left.\bar{p}^{0}_{\star}\right|_{\uvec{q}=\uvec{0}}= \sqrt{\bar{\uvec{p}}^{2} + M^{2}} = E_{\bar{\uvec{p}}}$ and $\left.q^{0}_{\star}\right|_{\uvec{q}=\uvec{0}}=0$. On the level of distributions Eq.~\eqref{P_deriv} implies that the matrix element $\overline{\mathcal{T}}$ can be explicitly written  
\begin{align}
\overline{\mathcal{T}}(\bar{p},q) = 2\pi \, \delta_{M}^{(+)}(\bar{p}) \frac{\overline{C}(\bar{p},0)}{2\bar{p}^{0}}   \, \delta^{4}(q).
\label{cov_dist_simple}
\end{align}

\noindent
The calculation of the rotation and boost generator matrix elements instead requires one to evaluate more complicated distributional relations of the form
\begin{align}
\mathcal{T}^{j}(p',p) = \delta_{M}^{(+)}(p')\delta_{M}^{(+)}(p)\, C(p',p) \, \frac{\partial}{\partial p_{j}}\delta^{3}(\uvec{p}' -\uvec{p} ),
\end{align} 
Performing an identical procedure as before, and applying the definition of the distributional derivative~\cite{Strichartz94}, the smeared distribution $\overline{\mathcal{T}}^{j}(\bar{p},q) = \mathcal{T}^{j}(p',p)$ is given by
\begin{align}
\int \ud^{3}\bar{\uvec{p}} \  \left. \frac{\partial}{\partial q_{j}} \! \left( \frac{\overline{C}(\bar{p}^{0}_{\star}, \, \bar{\uvec{p}}, \, q^{0}_{\star}, \, \uvec{q}) \,  f(\bar{p}^{0}_{\star}, \, \bar{\uvec{p}}, \, q^{0}_{\star}, \, \uvec{q})}{4\sqrt{(\bar{\uvec{p}} + \tfrac{1}{2}\uvec{q})^2 +M^{2}}\sqrt{(\bar{\uvec{p}} - \tfrac{1}{2}\uvec{q})^2 +M^{2}}} \right) \right|_{\uvec{q}=\uvec{0}}
\end{align}   
Differentiating the denominator and evaluating at $\uvec{q}=\uvec{0}$ leads to a vanishing expression, so the only terms which contribute are the derivatives of the coefficient and the test function. Since $\bar{p}^{0}$ and $q^{0}$ are set to $\bar{p}^{0}_{\star}$ and $q^{0}_{\star}$ respectively, both of which depend on $\uvec{q}$, this results in additional terms besides those that arise due to the explicit $\uvec{q}$-dependence of $f$ and $\overline{C}$. Besides the fact that $\left.\bar{p}^{0}_{\star}\right|_{\uvec{q}=\uvec{0}}=  E_{\bar{\uvec{p}}}$ and $\left.q^{0}_{\star}\right|_{\uvec{q}=\uvec{0}}=0$, it also follows from Eqs.~\eqref{eq_p0} and~\eqref{eq_q0} that 
\begin{align}
\left.\frac{\partial \bar{p}^{0}_{\star}}{\partial q_{j}}\right|_{\uvec{q}=\uvec{0}}=0, \hspace{10mm} \left.\frac{\partial q^{0}_{\star}}{\partial q_{j}}\right|_{\uvec{q}=\uvec{0}}= -\frac{\bar{p}^{j}}{E_{\bar{\uvec{p}}}}.
\end{align}
After applying the chain rule together with the above identities one obtains 
\begin{align}
&\int \frac{\ud^{3}\bar{\uvec{p}}}{(2E_{\bar{\uvec{p}}})^{2}} 
\Bigg[  \left\{ -\frac{\bar{p}^{j}}{E_{\bar{\uvec{p}}}}  \left.\frac{\partial \overline{C}(E_{\bar{\uvec{p}}}, \bar{\uvec{p}},q^{0}, \uvec{q})}{\partial q^{0}}\right|_{q^{0}=q^{0}_{\star}} + \frac{\partial \overline{C}(E_{\bar{\uvec{p}}}, \bar{\uvec{p}},0, \uvec{q})}{\partial q_{j}} \right\}    f(E_{\bar{\uvec{p}}}, \bar{\uvec{p}},0, \uvec{q}) \nonumber \\
& \hspace{25mm} -   \overline{C}(E_{\bar{\uvec{p}}}, \bar{\uvec{p}},0, \uvec{q})\left\{ -\frac{\bar{p}^{j}}{E_{\bar{\uvec{p}}}}  \left.\frac{\partial f(E_{\bar{\uvec{p}}}, \bar{\uvec{p}},q^{0}, \uvec{q})}{\partial q^{0}}\right|_{q^{0}=q^{0}_{\star}} + \frac{\partial f(E_{\bar{\uvec{p}}}, \bar{\uvec{p}},0, \uvec{q})}{\partial q_{j}} \right\}    \Bigg]_{\uvec{q}=\uvec{0}}, 
\end{align}
which on the level of distributions implies
\begin{align}
\overline{\mathcal{T}}^{j}(\bar{p},q) = -2\pi\frac{\delta_{M}^{(+)}(\bar{p})}{2\bar{p}^{0}} \left[ \overline{C}(\bar{p},0) \, \partial^{j}\delta^{4}(q) - \overline{C}(\bar{p},0) \, \frac{\bar{p}^{j}}{\bar{p}^{0}} \, \partial^{0}\delta^{4}(q) - \left(\frac{\partial \overline{C}}{\partial q_{j}} -\frac{\bar{p}^{j}}{\bar{p}^{0}} \frac{\partial \overline{C}}{\partial q^{0}}\right)_{q=0}  \delta^{4}(q)    \right]. 
\label{cov_dist}
\end{align}

\subsection{Explicit matrix elements}

In order to perform the distributional matching procedure one requires the explicit forms for the rotation and boost generator matrix elements. In the variables $p'$ and $p$ these are given by
\begin{align}
\langle p',m';M| J^{i}|p,m;M\rangle &= (2\pi)^{4}\delta_{M}^{(+)}(p)\left[\Sigma^{i}_{m'm}(k) + \delta_{m'm} \, i\epsilon^{ijk}p^{k}\frac{\partial}{\partial p_{j}}  \right]\delta^{4}(p'-p), \label{JM_1} \\
\langle p',m';M| K^{i}|p,m;M\rangle &= -(2\pi)^{4}\delta_{M}^{(+)}(p)\bigg[\frac{\epsilon^{ijk}p^{j}}{p^{0}+M}\, \Sigma^{k}_{m'm}(k)  \nonumber \\
& \hspace{25mm} + \delta_{m'm} \, i \left(p^{0} \frac{\partial}{\partial p_{i}} - p^{i} \frac{\partial}{\partial p_{0}} \right) \bigg]\delta^{4}(p'-p), \label{KM_1}
\end{align}
which are a covariant generalisation of those derived in~\cite{Bakker_Leader_Trueman04}. To derive these equations one can use the fact that states of arbitary spin $s$ transform under (proper orthochronous) Lorentz transformations $\alpha$ as follows~\cite{Haag96}:
\begin{align}
U(\alpha)|p,k;M\rangle = \sum_{l}\mathcal{D}^{(s)}_{lk}(\alpha)|\Lambda(\alpha)p,l;M\rangle,
\label{Lorentz_tran}
\end{align} 
where $\mathcal{D}^{(s)}$ is the $(2s+1)$-dimensional Wigner rotation matrix, and $\Lambda(\alpha)$ is the four-vector representation of $\alpha$. Since we are interested in the matrix elements of $J^{i}$ and $K^{i}$ one must consider the specific cases of a pure rotation $\alpha= \mathcal{R}_{i}$ and boost $\alpha= \mathcal{B}_{i}$ about the $i$-axis, where: $U(\mathcal{R}_{i}) = e^{-i \beta J^{i}}$ and $U(\mathcal{B}_{i}) = e^{i \xi K^{i}}$. Combining Eq.~\eqref{Lorentz_tran} for a pure rotation together with the definition of the norm of the on-shell states in Eq.~\eqref{norm} implies 
\begin{align}
\langle p',m';M| J^{i}|p,m;M\rangle &= i\left[\frac{\partial}{\partial \beta} \langle p',m';M|U(\mathcal{R}_{i})|p,m;M\rangle  \right]_{\beta=0}   \nonumber \\ 
&= i\left.\frac{\partial}{\partial \beta} \left(                                
   \sum_{l}\mathcal{D}^{(s)}_{lm}(\mathcal{R}_{i}) \, \langle p',m';M|\Lambda(\mathcal{R}_{i})p,l;M\rangle    \right)\right|_{\beta=0}  \nonumber \\
&= i\left.\frac{\partial}{\partial \beta} \left(                                
   \sum_{l}\mathcal{D}^{(s)}_{lm}(\mathcal{R}_{i})\,  (2\pi)^{4} \delta^{4}(p'-\Lambda(\mathcal{R}_{i})p) \, \delta_{M}^{(+)}(p') \delta_{m'l}   \right)\right|_{\beta=0} \nonumber \\
&= i \left[\frac{\partial}{\partial \beta} \mathcal{D}^{(s)}_{lm}(\mathcal{R}_{i})\right]_{\beta=0}(2\pi)^{4}\delta^{4}(p'-p) \, \delta_{M}^{(+)}(p)  \nonumber \\
& \hspace{20mm} +(2\pi)^{4}\delta_{M}^{(+)}(p) \, \delta_{m'm} \, i \left[\frac{\partial}{\partial \beta}   \delta^{4}(p'-\Lambda(\mathcal{R}_{i})p) \right]_{\beta=0}, 
\label{J_M1}
\end{align}  
where one has implicitly used the fact that $\delta_{M}^{(+)}(\Lambda(\mathcal{R}_{i})p) = \delta_{M}^{(+)}(p)$. By definition: $\Sigma^{i}_{m'm}(k)=i\left[\frac{\partial}{\partial \beta} \mathcal{D}^{(s)}_{m'm}(\mathcal{R}_{i})\right]_{\beta=0}$ are the $(2s+1)$-dimensional spin matrices. To consistently calculate the second term one must use the distributional properties of the Dirac delta. In general, due to the transformation properties of distributions under linear transformations~\cite{Strichartz94}, one has that
\begin{align}
\int \ud^{4}p \ \delta^{4}(p'-\Lambda(\mathcal{R}_{i})p) \, f(p) &\equiv |\text{det}\Lambda(\mathcal{R}_{i})|^{-1}\int \ud^{4}\ell \ \delta^{4}(p'-\ell) \, f(\Lambda^{-1}(\mathcal{R}_{i})\ell) \nonumber \\
&= f(\Lambda^{-1}(\mathcal{R}_{i})p'),
\label{linear}
\end{align} 
where $f$ is an arbitrary test function. Expanding the test function around the point $\beta=0$ gives
\begin{align}
f(\Lambda^{-1}(\mathcal{R}_{i})p') = f(p^{\prime}) + \beta \, \epsilon^{ijk} \, p^{\prime j}\left.\frac{\partial f(p)}{\partial p_{k}}\right|_{p=p'} + \mathcal{O}(\beta^{2}).
\end{align}  
Combining this expansion together with Eq.~\eqref{linear} one can then explicitly determine how the distribution $i\left[\frac{\partial}{\partial \beta}  \delta^{4}(p'-\Lambda(\mathcal{R}_{i})p) \right]_{\beta=0}$ acts on test functions
\begin{align}
\int \ud^{4}p \ i\left[\frac{\partial}{\partial \beta}  \delta^{4}(p'-\Lambda(\mathcal{R}_{i})p) \right]_{\beta=0}f(p) &= i \left.\frac{\partial}{\partial \beta} \left(  f(p^{\prime i}) + \beta \, \epsilon^{ijk} \, p^{\prime j}\left.\frac{\partial f(p)}{\partial p_{k}}\right|_{p=p'} + \mathcal{O}(\beta^{2}) \right)\right|_{\beta=0} \nonumber \\
&=  i\epsilon^{ijk}p^{ \prime j}\left.\frac{\partial f(p)}{\partial p_{k}}\right|_{p=p'},
\end{align}
which implies the following equality:
\begin{align}
i \left[\frac{\partial}{\partial \beta}   \delta^{4}(p'-\Lambda(\mathcal{R}_{i})p) \right]_{\beta=0} = i\epsilon^{ijk}p^{k}\frac{\partial}{\partial p_{j}}\delta^{4}(p'-p).
\label{J_M2}
\end{align}
Combining this relation with Eq.~\eqref{J_M1} finally proves Eq.~\eqref{JM_1}. \\

\noindent
In the case of a pure boost $\alpha= \mathcal{B}_{i}$ the matrix element is more complicated because the Wigner rotation matrix $\mathcal{D}^{(s)}(\mathcal{B}_{i})$ depends on both $\xi$ and the momentum. Nevertheless, one can demonstrate that\footnote{In~\cite{Bakker_Leader_Trueman04} the authors derive the form for the infinitesimal Wigner rotation for boosts, from which one can derive the manifestly spin-representation independent expression in Eq.~\eqref{Wigner_boost}.}  
\begin{align}
i \left[\frac{\partial}{\partial \xi} \mathcal{D}^{(s)}_{m'm}(\mathcal{B}_{i})\right]_{\xi=0} = \frac{\epsilon^{ijk}p^{j}}{p^{0}+M}\Sigma^{k}_{m'm}(k).
\label{Wigner_boost}
\end{align}  
Performing identical steps as in Eq.~\eqref{J_M1}, it remains to calculate an explicit expression for the distribution $i\left[\frac{\partial}{\partial \xi}   \delta^{4}(p'-\Lambda(\mathcal{B}_{i})p) \right]_{\xi=0}$. In this case
\begin{align}
f(\Lambda^{-1}(\mathcal{B}_{i})p') = f(p^{\prime})  - \xi \, p^{\prime 0}\left.\frac{\partial f(p)}{\partial p_{i}}\right|_{p=p'} + \xi \, p^{\prime i}\left.\frac{\partial f(p)}{\partial p_{0}}\right|_{p=p'} + \mathcal{O}(\xi^{2}),
\end{align} 
from which it follows
\begin{align}
i \left[\frac{\partial}{\partial \xi}   \delta^{4}(p'-\Lambda(\mathcal{B}_{i})p) \right]_{\xi=0} = \left[ ip^{0}\frac{\partial}{\partial p_{i}} - ip^{i}\frac{\partial}{\partial p_{0}} \right]\delta^{4}(p'-p).
\label{K_M2}
\end{align}
Combining this with Eq.~\eqref{Wigner_boost} proves Eq.~\eqref{KM_1}. \\

\noindent
In order to compare these equations with the on-shell matrix elements one must instead work with the variables $\bar{p}$ and $q$. Due to the explicit $\delta^{4}(p'-p)$ component in the first terms of Eqs.~\eqref{JM_1} and~\eqref{KM_1}, these expressions are simply proportional to $\delta_{M}^{(+)}(\bar{p})\delta^{4}(q)$. The second terms involving derivatives of $\delta^{4}(p'-p)$ are non-trivial though due to the $q$-dependence of $\delta_{M}^{(+)}(p)$. Nevertheless, in the case of rotations one can write
\begin{align}
\delta_{M}^{(+)}(p) \, i\left[ \frac{\partial}{\partial \beta}  \delta^{4}(p'-\Lambda(\mathcal{R}_{i})p) \right]_{\beta=0} &=  -\frac{\delta\left(\bar{p}^{0}-\tfrac{1}{2}q^{0} - \sqrt{(\bar{\uvec{p}}-\tfrac{1}{2}\uvec{q})^{2} + M^{2}} \right)}{2(\bar{p}^{0}-\tfrac{1}{2}q^{0})} i\epsilon^{ijk}(\bar{p}-\tfrac{1}{2}q)^{k}\frac{\partial}{\partial q_{j}}\delta^{4}(q) \nonumber \\
&=  -\frac{\delta\left(\bar{p}^{0} - \sqrt{(\bar{\uvec{p}}-\tfrac{1}{2}\uvec{q})^{2} + M^{2}} \right)}{2\bar{p}^{0}} i\epsilon^{ijk}\bar{p}^{k}\frac{\partial}{\partial q_{j}}\delta^{4}(q), 
\end{align}   
since the term involving the $q_{j}$-derivative of $q^{k}$ vanishes due to the anti-symmetric tensor. If one now integrates this expression with a test function $f(\bar{p},q)$ one ends up with
\begin{align}
\int \ud^{3}\bar{p} \ i\epsilon^{ijk}\bar{p}^{k} \left[ -\frac{\bar{p}^{j}}{4E_{\bar{\uvec{p}}}^{3}}f(E_{\bar{\uvec{p}}}, \bar{\uvec{p}},q_{0}, \uvec{q}) +\frac{\bar{p}^{j}}{4E_{\bar{\uvec{p}}}^{2}} \left.\frac{\partial f(\bar{p}_{0}, \bar{\uvec{p}},q_{0}, \uvec{q})}{\partial \bar{p}_{0}}\right|_{\bar{p}_{0}=2E_{\bar{\uvec{p}}}} + \frac{1}{2E_{\bar{\uvec{p}}}}\frac{\partial f(E_{\bar{\uvec{p}}}, \bar{\uvec{p}},q_{0}, \uvec{q})}{\partial q_{j}} \right]_{q=0}.
\end{align} 
The first two terms vanish due to the contraction with $\epsilon^{ijk}\bar{p}^{k}$, and hence one can conclude that
\begin{align}
\delta_{M}^{(+)}(p) \, i\epsilon^{ijk}p^{k}\frac{\partial}{\partial p_{j}}\delta^{4}(p'-p) =  -\delta_{M}^{(+)}(\bar{p}) \, i\epsilon^{ijk}\bar{p}^{k}\frac{\partial}{\partial q_{j}}\delta^{4}(q).
\end{align}
The $J^{i}$ matrix element in $(\bar{p},q)$ variables is therefore given by
\begin{align}
\langle \bar{p}+ \tfrac{1}{2}q,m';M| J^{i}|\bar{p}- \tfrac{1}{2}q,m;M\rangle &= (2\pi)^{4}\delta_{M}^{(+)}(\bar{p})\left[\Sigma^{i}_{m'm}(k) - \delta_{m'm} \, i\epsilon^{ijk}\bar{p}^{k}\frac{\partial}{\partial q_{j}}  \right]\delta^{4}(q). 
\label{JM_pbarQ} 
\end{align}
\ \\
\noindent
One can perform exactly the same procedure in the pure boost case, except this time there are two derivative components. Changing variables in the expression $\delta_{M}^{(+)}(p)\left[ ip^{0}\frac{\partial}{\partial p_{i}} - ip^{i}\frac{\partial}{\partial p_{0}} \right]\delta^{4}(p'-p)$ and integrating with a test function gives
\begin{align}
i\int \ud^{3}\bar{p} \ \left[ \frac{1}{2} \frac{\partial f(E_{\bar{\uvec{p}}}, \bar{p},q_{0}, \uvec{q})}{\partial q_{i}} - \frac{\bar{p}^{i}}{2E_{\bar{\uvec{p}}}}\frac{\partial f(E_{\bar{\uvec{p}}}, \bar{p},q_{0}, \uvec{q})}{\partial q_{0}} \right]_{q=0},
\end{align}
where the two terms involving $\bar{p}_{0}$-derivatives of the test function cancel one another. From this we conclude that
\begin{align}
\delta_{M}^{(+)}(p)\left[ ip^{0}\frac{\partial}{\partial p_{i}} - ip^{i}\frac{\partial}{\partial p_{0}} \right]\delta^{4}(p'-p) = -\delta_{M}^{(+)}(\bar{p})\left[ i\bar{p}^{0}\frac{\partial}{\partial q_{i}} - i\bar{p}^{i}\frac{\partial}{\partial q_{0}} \right]\delta^{4}(q),
\end{align} 
and hence the $K^{i}$ matrix element in $(\bar{p},q)$ variables has the form
\begin{align}
\langle \bar{p}+ \tfrac{1}{2}q,m';M| K^{i}|\bar{p}- \tfrac{1}{2}q,m;M\rangle &= (2\pi)^{4}\delta_{M}^{(+)}(\bar{p})\bigg[-\frac{\epsilon^{ijk}\bar{p}^{j}}{\bar{p}^{0}+M}\, \Sigma^{k}_{m'm}(k)  \nonumber \\
& \hspace{25mm} + \delta_{m'm} \, i \left(\bar{p}^{0} \frac{\partial}{\partial q_{i}} - \bar{p}^{i} \frac{\partial}{\partial q_{0}} \right) \bigg]\delta^{4}(q).
\label{KM_pbarQ} 
\end{align}

\section{Arbitrary spin $\eta_{m}(p)$ identities}
\label{appendix_b}

In this appendix we prove a series of identities involving the arbitrary spin $\eta_{m}(p)$.

\subsection{Proof of Eq.~\eqref{rel}} 

In order to prove Eq.~\eqref{rel} it is important to first recognise that one can write
\begin{align}
\frac{\partial}{\partial q_{i}}\left[\overline{\eta}_{m'}(p')\eta_{m}(p)\right] &= \left[\frac{\partial \overline{\eta}_{m'}(\bar{p}+\tfrac{1}{2}q)}{\partial q_{i}} \right]\eta_{m}(\bar{p}) + \overline{\eta}_{m'}(\bar{p}) \left[\frac{\partial \eta_{m}(\bar{p}-\tfrac{1}{2}q)}{\partial q_{i}} \right]\nonumber \\
&= \left[\frac{\partial \overline{\eta}_{m'}(\bar{p}+\tfrac{1}{2}q)}{\partial q_{i}} \right]\eta_{m}(\bar{p}) - \overline{\eta}_{m'}(\bar{p}) \left[\frac{\partial \eta_{m}(\bar{p}+\tfrac{1}{2}q)}{\partial q_{i}} \right].
\label{eta_deriv}
\end{align}
The rest and moving frame $\eta_{m}$ are related by a global boost: $\eta_{m}(p) = e^{i\uvec{\xi}(p)\cdot \uvec{\kappa}}\,\eta_{m}(k)$, where $k^{\mu} = Mg^{\mu 0}$ is the rest frame four-momentum, $\kappa^i=S^{0i}$ are the standard boost generator matrices in the chosen spin representation, and the boost parameter is given by: $\uvec\xi(p) = \xi(p) \,\hat{\uvec \xi}(p)$, with $\xi(p)= \sinh^{-1}(|\uvec{p}|/M)$ and $\hat{\uvec \xi}(p)=\uvec p/|\uvec p|$. Let us first consider the derivative of the exponential argument in this boost, evaluated at $q=0$
\begin{align}
\left.\frac{\partial}{\partial q_{i}}\!\left[i\uvec{\xi}(\bar{p}+\tfrac{1}{2}q)\cdot \uvec{\kappa}\right]\right|_{q=0} &= \left[\frac{\partial}{\partial q_{i}}\xi(\bar{p}+\tfrac{1}{2}q)\right]_{q=0}(i \hat{\uvec{\xi}}\cdot \uvec{\kappa}) + \xi \left.\frac{\partial}{\partial q_{i}}\!\left[i\hat{\uvec{\xi}}(\bar{p}+\tfrac{1}{2}q)\cdot \uvec{\kappa}\right]\right|_{q=0} \nonumber \\
&= -\frac{\bar{p}^{i}}{2|\bar{\uvec{p}}|\bar{p}^{0}\xi}\,  (i \uvec{\xi}\cdot \uvec{\kappa}) + \frac{i}{2|\bar{\uvec{p}}|}\,\epsilon^{ijk}\epsilon^{klr}\hat{\xi}^{j}\xi^{l}\kappa^{r} \nonumber \\
&= -\frac{\bar{p}^{i}}{2|\bar{\uvec{p}}|\bar{p}^{0}\xi}\,  (i \uvec{\xi}\cdot \uvec{\kappa}) - \frac{i}{2|\bar{\uvec{p}}|^2}\,\epsilon^{ijk}\bar p^{j}\left[\Sigma^{k}, (i\uvec{\xi}\cdot \uvec{\kappa})  \right],
\end{align} 
where we used the fact that the boost generators transform as a three-vector under rotations $[\Sigma^k,\kappa^l]=i\epsilon^{klr}\kappa^r$. Because the commutator with $\Sigma^{k}=\frac{1}{2}\epsilon^{kij}S^{ij}$ acts as a derivation, it follows from the above relation that the $q_{i}$-derivative on the full exponential can be written
\begin{align}
\left[\frac{\partial}{\partial q_{i}}e^{i\uvec{\xi}(\bar{p}+\tfrac{1}{2}q)\cdot \uvec{\kappa}}\right]_{q=0} = -\frac{\bar{p}^{i}}{2|\bar{\uvec{p}}|\bar{p}^{0}\xi}\,(i \uvec{\xi}\cdot \uvec{\kappa}) 
\, e^{i\uvec{\xi}(\bar{p}) \cdot \uvec{\kappa}} - \frac{i}{2|\bar{\uvec{p}}|^{2}}\,\epsilon^{ijk}\bar{p}^{j}\left[\Sigma^{k}, e^{i\uvec{\xi}(\bar{p})\cdot \uvec{\kappa}}  \right],
\end{align}
and similarly with $\uvec\kappa\mapsto -\uvec\kappa$. Using these expressions together with Eq.~\eqref{eta_deriv} one finds that
\begin{align}
\left.\frac{\partial}{\partial q_{i}}\left[\overline{\eta}_{m'}(p')\eta_{m}(p)\right]\right|_{q=0} &= \text{Tr}\left[\rho_{m'm}(k) \, \left[\frac{\partial}{\partial q_{i}}e^{-i\uvec{\xi}(\bar{p}+\tfrac{1}{2}q)\cdot \uvec{\kappa}}\right]_{q=0} e^{i\uvec{\xi}(\bar{p}) \cdot \uvec{\kappa}}    \right]   \nonumber  \\
& \quad - \text{Tr}\left[\rho_{m'm}(k) \,  e^{-i\uvec{\xi}(\bar{p}) \cdot \uvec{\kappa}} \left[\frac{\partial}{\partial q_{i}}e^{i\uvec{\xi}(\bar{p}+\tfrac{1}{2}q)\cdot \uvec{\kappa}}\right]_{q=0}   \right]  \nonumber \\
&=\frac{\bar{p}^{i}}{|\bar{\uvec{p}}|\bar{p}^{0}\xi}\, \text{Tr}\left[\rho_{m'm}(k) \,(i \uvec{\xi}\cdot \uvec{\kappa})  \right]  \nonumber  \\
& \quad +\frac{i}{|\bar{\uvec{p}}|^{2}}\,\epsilon^{ijk}\bar{p}^{j} \,  \text{Tr} \left[\rho_{m'm}(k)\left\{ \left(e^{-i\uvec{\xi}(\bar{p})\cdot \uvec{\kappa}}\Sigma^k e^{i\uvec{\xi}(\bar{p})\cdot \uvec{\kappa}} \right) - \Sigma^k  \right\} \right].
\label{combine}
\end{align}
The first term vanishes because of the trace\footnote{This trace must indeed to vanish, otherwise a state at rest would be characterised by an additional three-vector besides the spin vector.} $\text{Tr}\left[\rho_{m'm}(k)\kappa^i  \right]=0$, and one is left with
\begin{align}
\left.\frac{\partial}{\partial q_{i}}\left[\overline{\eta}_{m'}(p')\eta_{m}(p)\right]\right|_{q=0} = \frac{i}{|\bar{\uvec{p}}|^{2}}\,\epsilon^{ijk}\bar{p}^{j} \left\{  \text{Tr} \left[\rho_{m'm}(\bar{p})\Sigma^{k}\right] -\text{Tr} \left[\rho_{m'm}(k)\Sigma^{k}\right]   \right\}.
\end{align}

\subsection{Proof of Eq.~\eqref{J_contract}} 

In order to prove Eq.~\eqref{J_contract} note that since a state is characterised only in terms of the momentum and Pauli-Lubanski four-vectors, one can in general write~\cite{Lorce:2018zpf}
\begin{align}
\overline\eta_{m'}(\bar p)S^{\mu\nu}\eta_m(\bar p)=-\frac{1}{M^2}\,\epsilon^{\mu\nu}_{\phantom{\mu\nu}\alpha\beta}\,S^\alpha_{m'm}(\bar p)\,\bar p^\beta,
\label{Lorentzidentity}
\end{align}
where $S^\alpha_{m'm}(\bar{p})=\text{Tr}\left[\rho_{m'm}(\bar p)S^\alpha\right]$ with $S^{\alpha}=\frac{1}{2}\epsilon^\alpha_{\phantom{\mu}\rho\sigma\lambda}S^{\rho\sigma}\bar p^\lambda$ the standard covariant spin matrices in the chosen spin representation, and $\epsilon_{0123}=+1$. Contracting this relation with the four-momentum leads to $\overline\eta_{m'}(\bar p)S^{\mu\nu}\eta_m(\bar p)\bar p_{\mu}=0$, and hence
\begin{align}
\epsilon^{ijk}\bar{p}^{j}\,\Sigma^k_{m'm}(\bar p)=-\bar p^0\,\kappa^i_{m'm}(\bar p),
\label{id1}
\end{align}
since $\kappa^i=S^{0i}$ and $\Sigma^k=\frac{1}{2}\epsilon^{kij} S^{ij}$. Another consequence of Eq.~\eqref{Lorentzidentity} is that
\begin{align}
\overline\eta_{m'}(\bar p)S^{\mu\nu}\eta_m(\bar p)k_{\mu}   &=\frac{1}{M^2}\,\epsilon^{\nu}_{\phantom{\mu}\mu\alpha\beta}k^{\mu} S^\alpha_{m'm}(\bar p)\,\bar p^\beta \nonumber \\
&= -\frac{1}{M^2}\,\epsilon^{\nu}_{\phantom{\mu}\mu\alpha\beta}k^{\mu} S^\alpha_{m'm}(k)\,\bar p^\beta =\overline\eta_{m'}(k)S^{\mu\nu}\eta_m(k)\bar{p}_{\mu},
\end{align}
and hence it follows that
\begin{align}
M\kappa^i_{m'm}(\bar p)=-\epsilon^{ijk}\overline p^j\,\Sigma^k_{m'm}(k).
\label{id2}
\end{align}
Combining Eqs.~\eqref{id1} and~\eqref{id2} together with $(\bar p^0)^2=|\bar{\uvec p}|^2+M^2$ leads us to
\begin{align*}
\epsilon^{ijk}\bar{p}^{j}\bar{p}^{0}\, \Sigma^k_{m'm}(\bar{p})  = -|\bar{\uvec{p}}|^{2}\, \kappa^i_{m'm}(\bar{p})  + M \epsilon^{ijk}\bar{p}^{j}\, \Sigma^k_{m'm}(k).
\end{align*} 

\bibliographystyle{JHEP}

\bibliography{paper_refs}

\end{document}